\documentclass[]{article}
\usepackage[latin1]{inputenc}
\usepackage[english]{babel}
\usepackage{amsmath}
\usepackage{amsfonts}
\usepackage{amssymb}
\usepackage{url}
\usepackage{color}
\usepackage{listings}
\usepackage[numbers,sort&compress]{natbib}

\newcommand\csharp{C$^\#$}

\definecolor{light-gray}{gray}{0.95}
\begin{document}

\title{Array Requirements for Scientific Applications and
an Implementation for Microsoft SQL Server}

\author{L\'aszl\'o Dobos$^{1,2}$, Alexander Szalay$^{2}$, Jos\'e Blakeley$^{3}$ \\ Tam\'as Budav\'ari$^{2}$, Istv\'an Csabai$^{1,2}$ \\ Dragan Tomic$^{3}$, Milos Milovanovic$^{3}$ \\ Marko Tintor$^{3}$ and Andrija Jovanovic$^{3}$}

\footnotetext[1]{E\"otv\"os Lor\'and University, Department of Physics of Complex Systems, 1117 Budapest, Hungary}
\footnotetext[2]{The Johns Hopkins University, Department of Physics and Astronomy, Baltimore, MD 21218, USA}
\footnotetext[3]{Microsoft Corporation}

\maketitle

\begin{abstract}

This paper outlines certain scenarios from the fields of astrophysics and fluid dynamics simulations which require high performance data warehouses that support array data type. A common feature of all these use cases is that subsetting and preprocessing the data on the server side (as far as possible inside the database server process) is necessary to avoid the client-server overhead and to minimize IO utilization. Analyzing and summarizing the requirements of the various fields help software engineers to come up with a comprehensive design of an array extension to relational database systems that covers a wide range of scientific applications.
We also present a working implementation of an array data type for Microsoft SQL Server 2008 to support large-scale scientific applications. We introduce the design of the array type, results from a performance evaluation, and discuss the lessons learned from this implementation. The library can be downloaded from our website at \url{http://voservices.net/sqlarray/}.

\end{abstract}

\paragraph*{Categories and Subject Descriptors}
H.2.4 [Database Management]: Systems -- relational databases; H.3.2 [Information Storage and Retrieval] Information storage; H.2.8 [Database Management] Database applications -- statistical databases;
E.1 [Data structures] -- Arrays;
J.2 [Physical Sciences and Engineering] -- Mathematics and statistics

\paragraph*{General Terms}
Performance, Design, Languages

\paragraph*{Keywords}
Database Server Programming, Extensibility, Abstract Data Types, Array Types

\section{Introduction}

With the increased need to execute complex calculations close to the data inside the database, it is becoming important to perform computations on arrays of data. These arrays often interact with standard libraries, like LAPACK, FFTW, SAS, R, MATLAB, and IDL. Long-running numerical simulations of computational fluid dynamics generate large, regularly spaced grids of data. For astronomy, high-throughput genomics, and fluid dynamics -- applications we are considering -- it is not enough to marshal blocks of data back and forth between the database and the libraries. We also need to perform various array manipulations, like extracting subsets, or computing aggregates over certain dimensions. Also, some of these scenarios require storing data of multi-dimensional parameter spaces on a point-by-point basis where the individual arrays are small in size but the number of rows is high (in the 109 regime). Efficient search in these multi-dimensional datasets is also an important objective.

Over the last two decades there was a lot of work realized on arrays in databases \cite{DBLP:conf/pods/MaierV93, DBLP:conf/ngits/Baumann99, rasdaman, DBLP:conf/dexa/BallegooijCVK05, Ballegooij03ram:array}. The open source Postgres database also contains an array data type, with a fairly extensive language binding to SQL \cite{postgresql}. Over the last two years the SCIDB project is involved in building a very scalable database where the primitive data type is a multidimensional array \cite{DBLP:conf/cidr/StonebrakerBDLMRZ09, DBLP:conf/sigmod/Brown10}.

While it is certainly possible to build various simple array types as UDTs in the application, having a well-defined, common, built-in array type in the DBMS provides better performance and a consistent API that external libraries could be interfaced with, enabling applications to build on top of one another. The alternative of each user builds their own array type results in a smaller and more fragmented collection of tools. 

Below we list our requirements for the array library to be useful for the scientific databases we are using today:

\begin{itemize}

\item Have a T-SQL interface, and appear as a new set of data types
\item Interface supports multidimensional arrays (up to at least 6 dimensions) of variable length
\item The major base data types are supported (bigint, int, smallint, tinyint, real, float, datetime, complex)
\item Simple way to create an array of a given size
\item Arrays are directly compatible with LAPACK, MATLAB, FFTW etc.
\item Simple T-SQL interface to access the dimensions/sizes of a given array
\item Simple T-SQL interface to extract various (possibly lower dimensional) subsets
\item Simple T-SQL interface to perform various aggregate operations over arrays
\item Recast the dimensions of an array (while keeping the size fixed)
\item Efficient support for both small (in-page) and large (out-of-page) arrays

\end{itemize}

In Sec.~\ref{sec:science} we describe the scientific use-cases that led us to design and build a library that adds an array type to Microsoft SQL Server. The prototype implementation details are provided in Sec.~\ref{sec:prototype}. Challenges of the implementation are described in Sec.~\ref{sec:implementation}, the T-SQL and .NET interface is briefly introduced in Sec.~\ref{sec:interface}, and initial performance test results are summarized in Sec.~\ref{sec:performance}. Lessons learned from the implementation are discussed in Sec.~\ref{sec:discussion} and \ref{sec:conclusions}.

\section{Scientific use cases}
\label{sec:science}

In this section, we introduce the scenarios that provided the motivation for this work. Our philosophy about database-centric computing was described in detail in a volume dedicated to our collaborator, Jim Gray \cite{Szalay2008}. Here we present a few science scenarios, mostly drawn from physical sciences, which show how beneficial the integration of monolithic arrays of a primitive data type and a SQL database can be. 

While most of the scientific simulations compute quantities on a regular grid and store results in large, dense arrays, many scientific databases store multidimensional data not on a grid but on a point-by-point basis. This approach is very typical in astrophysical datasets where measurements of individual objects occupy a high-dimensional space (usually above five dimensions, but spectroscopic data can have dimensions of thousands). Search in these databases require special techniques not typically found in systems supporting geographical (thus predominantly at most 3D) data \cite{2007AN....328..852C, 2008ASPC..394..389D}.

\subsection{Simulations of Turbulent Flows}

Three years ago we built a prototype Microsoft SQL Server database that contains $2{,}000$ time steps from a $10{,}243$ simulation of a box with isotropic turbulence with a Reynolds number of 470 \cite{Li2008}. The simulation output is over a regular grid where every point contains the three components of the fluid velocity and the pressure. The data is partitioned along a space filling curve (z-index) into cubes of $(64+8)^3$. The $+8$ means that each cube contains an extra 8 voxel wide buffer so that particles on the edge of the original cube still have their neighbors within 4 voxels in the same blob. Each blob is about 6~MB and stored in a separate row.

The data is accessed via a web service where users can submit a set of about 10,000 particle positions and times and then can retrieve the interpolated values of the velocity field at those positions. This can be considered as the equivalent of placing small sensors into the simulation instead of downloading all the data or significant subsets of it. This service is public and is typically delivering about 108 particles per day world-wide (see \url{http://turbulence.pha.jhu.edu/}). Several papers appearing in the most appreciated physics journals (Phys. Rev. Letters, etc.) have used this facility. Currently we are adding a 70~TB simulation of a magneto-hydrodynamic system and next year we will add a 50~TB simulation of a channel flow. We are currently experimenting with different blob sizes, overlap regions and partitioning schemes across servers. Visualization services are around the corner and we are also considering enabling users to easily grab a sub-domain of the data.

The interpolation method provided by the service can be chosen from nearest point, PCHIP, and 4-6-8 point Lagrangian interpolation schemes. For the 8 point interpolation we need to convolve an 83 neighborhood with an 83 interpolation kernel for each point. Accessing the whole blob (6~MB) for an 8-point 3D interpolation is obviously overkill. By using much smaller blobs, especially if they fit onto a single 8~kB page, we could have a much lower overhead on disk IOs. If those are still appropriately clustered along a space filling curve, even disk access could be controller at the application level. 

\subsection{Astronomical Spectrum Databases}

A generic vector library could also be useful for storing and processing astronomical spectra. Spectra are measurements of the electromagnetic flux of astronomical objects in numerous energy (wavelength) bins and represented as a number of vectors such as wavelength bins (min, max and center wavelength), flux, error of the measured flux and flags. Latter is usually a vector of 8 or 16 bit integers. As the wavelength scale can change from observation to observation due to various reasons and the scale is usually not linear it is necessary to store the wavelength vector of each spectrum separately. Spectra can be one, two or three dimensional. One-dimensional spectra are measured by integrating the electromagnetic flux coming from different areas of the object so the result will be a one dimensional vector of the flux. Two dimensional spectra are measured by using a slit: different fluxes are measured depending on the position along this slit. The slit is usually positioned over the diameter of extended celestial sources such as galaxies and the position along the slit refers to an angular radius on the sky. Storing two dimensional spectra requires two axis vectors: wavelength and position, and a two dimensional array of the flux. Also, error and flag arrays need to be two dimensional. Three dimensional spectra are measured using so called integral field spectrographs where a high number of optical fibers are organized in closely packed bundles which can cover the extended celestial sources entirely. This technique allows astronomers to measure the energy distribution of the electromagnetic radiation of the sources pixel by pixel. The resulting data cube will be three dimensional with one wavelength axis and two position axes.
The typical processing steps of spectra include the following. Normalization of the flux vector which requires integration of the flux in given wavelength ranges and multiplication by scalar. Certain corrections of physical effects require multiplying the flux vector with a number that is a function of the wavelength. Resampling the spectra to a common wavelength grid is also very important. These operations require functions using two arrays of different data types (double for the wavelength and flux, integer for the flags). In case of astronomical spectra the resampling should be done such a way that the integrated flux in any wavelength range remains the same. Different processing steps however might require resampling using higher order functions or special kernels. Once resampled to common grid, spectra can be averaged to get composites with high signal to noise ratio. Generic resampling and integration functions would be very useful that could run in the query processing loop of Microsoft SQL Server to do this processing on the fly. The averaging could be very easily solved using an aggregate function. Latter would allow us to group spectra by certain parameters (for example redshift of the observed galaxies) so composite spectra of objects at different cosmological distances could be computed with a simple SQL query.

Higher dimensional spectrum processing would require subsetting arrays and summation over certain axes to get, for example, the overall spectrum of an object that was originally observed with an integral field spectrograph.

Using principal component analysis (PCA) for spectrum classification is a widespread technique. Running PCA over a set of spectra requires resampling and normalization of the individual data vectors, computing the correlation matrix and executing a singular value decomposition (SVD) algorithm over the correlation matrix. The spectra then have to be expanded on the basis derived from the SVD which requires dot products between the data vectors and basis vectors. In practice, because of the flags that mask out wrong measurements bin by bin, dot product cannot be used for expanding spectra on a basis but least squares fitting is necessary which is again a very generic functionality that would be required in a vector library addressing a wide range of users. Certain spectrum processing operations also require non-negative least squares fitting.

When all spectra are expanded over a given orthogonal basis and coefficients are stored in a data column as a vector, similar spectrum search can be conducted the following way: One builds a kd-tree over the coefficients so nearest neighbor searches can be executed very quickly. A ``query'' spectrum is expanded on the same basis on the fly and the nearest neighbors of its coefficient vector are looked up using the kd-tree \cite{2008ASPC..394..389D}.

We developed Spectrum Services for the Virtual Observatory which already has a prototype of the vector data type implemented, though it can only handle one dimensional arrays and the implementation is purely client side and cannot leverage from the processing power of the machine running the database server. All the functionalities described above are also implemented but the current code is very specific to the data itself (mostly implemented inside a Spectrum class). Generic versions of the mentioned functions however would be easy to implement by reusing, modifying the code already written for Spectrum Services \cite{2007AN....328..852C, 2008ASPC..394..389D}.

\subsection{Cosmological N-body Simulations}

By Fall 2011 we will run 500 cosmological N-body simulations of 3203 particles each, with 100 snapshots each, dumping the ID, position and velocity for each particle, and a hash bucket ID, a time step, and simulation ID. This is a total of 40 bytes per point per snapshot amounting to $1.3$~GB per snapshot. The total volume of the simulations is expected to be about 66~TB. At around every $1{,}000$ time steps we will also output the Fourier transform of the density field on large scales which is a $100^3$ complex cube. The volume of this latter is small compared to the detailed snapshots.

It does not seem feasible to store the particle data broken down into individual rows. For the 100 snapshots and 500 simulations this would mean $1.6$ trillion rows. We need to arrange the data in coherent chunks organized into a spatial octree, not necessarily balanced \cite{2006astro.ph..8019L}. The octree would be computed from a space filling curve index. If we group together and store an order of a few thousand particles per bucket we can reduce the number of data table rows to about a billion but retrieving information about individual particles will require array-based data access.

Several analyses and operations will need to be performed on this data. At each snapshot we need to compute the so-called halos, clusters of particles identified by friends of friends (FOF) algorithms within a certain distance. This requires a lot of parallel neighbor calculations. These FOF halos need to be linked up between the different time steps to determine the so called merger history. This can be best done by comparing the particle labels in the halos at different time steps. A decimated octree of particles for several hierarchical levels also needs to be computed for the purposes of visualization where each sub-sampled particle would get a different weight according to the number of original particles in its region of attraction \cite{2006astro.ph..8019L}.

We will also need to compute the density over a 6403 grid, interpolating over the particle positions, using a cloud-in-cell (CIC) algorithm, then Fourier transform it and compute its power spectrum.

We will need to build light-cones through the simulations where we look at the cube from a distant viewpoint and follow light rays back into the simulation and recreate the galaxy velocities in an expanding universe including the Doppler-shift of the galaxies along the radial direction due to their velocities. Furthermore, as we look farther, the simulation box needs to be taken from an earlier time step since the light coming to us was emitted by those galaxies at a much earlier epoch. This requires a spatial index that can retrieve points from within a cone or other geometric primitives. Finally, we need to be able to compute various statistical functions like two and three point correlations over these point sets with distances calculated in the curved geometry of the universe.

\section{Implementation}
\label{sec:prototype}

\subsection{Main Objectives}

We have been using Microsoft SQL Server since version 2000 to build scientific data warehouses and experiencing the problem of lack of a native array data type. Some of our recent SQL Server databases containing data from turbulence computations are in the 30--50~TB regime, and rapidly growing \cite{Li2008}.

Since version 2005 \cite{DBLP:conf/sigmod/AchesonBBCEFJKRSSVZ04, DBLP:conf/sigmod/BlakeleyRKPHK08} .NET common language runtime (CLR) integration enables users to implement user code that runs inside the server process and eliminates the communication client-server overhead. Although developing a generic array framework in CLR has its limitations as we will discuss later, implementing arrays natively on the SQL Server codebase was beyond our possibilities so we decided to write the library in .NET.

The functions of the library were designed in such a way that calling them directly from T-SQL should be easy. As most client code to our scientific data warehouses is written in \csharp it was also an important objective to allow easy transition between our internal binary array format and standard .NET arrays.

We will discuss the limitations of SQL Server in reference with CLR implementation of a generic array data type below. Besides these limitations our aim was to deliver very good performance for small array (smaller than 8~kB in size) and reasonable performance with the most possible I/O optimizations for larger arrays.

\subsection{User-Defined Types in SQL Server CLR}

SQL Server supports executing user-defined stored procedures, scalar and table-valued functions, user-defined types (UDTs), and aggregate function (UDA) written in any .NET language compiled into .NET IL (intermediate language) code which runs on the .NET CLR (common language runtime). A known shortcoming of UDTs is, however, the lack of array support: only structures of basic data types can be used as UDTs unless one implements their own data serialization functions to read and write data into/from a binary data stream. The significant performance overhead of the data stream wrapper can be prohibitive so we decided to store array data as a binary data type and manipulate the blobs using plain functions instead of UDTs.

\subsection{The Two Different Storage Classes}
\label{sec:storage}

SQL Server treats binary objects differently depending on their size. Blobs smaller than 8~kB are stored on-page, as they fit into the 8~kB storage engine data pages. Blobs larger than 8~kB are stored out-of-page as B-trees. Access to out-of-page data is significantly slower than on-page data because (a) traversing B-trees is more expensive than simply addressing on-page data, and (b) out-of-page data has to go through the previously mentioned .NET binary stream wrapper that interfaces with the B-trees and provides random access to the blobs.

In order to gain the best performance for at least on-page arrays we implemented two storage classes based on array size. Arrays that fit into the data pages of the server are called short arrays while bigger data are called max arrays analogous to the \texttt{VARBINARY(MAX)} data type of SQL Server.

Short arrays are stored on-page in fixed-sized binary columns and can be converted to .NET arrays by a simple memory copy operation as they are available as \texttt{byte[]} buffers from SQL CLR. On the other hand, max arrays have to be read via the binary stream wrapper which has one important benefit: it supports reading only parts of the binary data if the whole array is not required. The latter can significantly speed up certain array subsetting operations. Short arrays have the limit of only six indices and indices are \texttt{Int16} while max arrays can have an unlimited number of indices and the index type is \texttt{Int32}.

\subsection{Data Type Support}

The current version of the array library only supports numeric data types: \texttt{Int8}, \texttt{Int16}, \texttt{Int32} and \texttt{Int64} (only signed), \texttt{float} and \texttt{double}. We do not support fixed-precision numbers as the main application of our library is for scientific data. For this reason we added support for float and double complex numbers as well. Scalar complex numbers are implemented as user-defined types and use the native serialization format of SQL Server.

\subsection{Data Format}

The arrays are stored as plain binary blobs decorated with a very simple header. In case of short arrays the header is 24 bytes long. We have flags to identify the type (short or max) and the underlying data type of the array so we can detect type mismatches at runtime when the blobs are passed to the wrong functions. The number of dimensions, the number of all elements and the sizes of the dimensions (up to six in case of short arrays or any number in case of max arrays) are also stored in the header. Because max arrays support any number of dimensions the header size may vary. Following the header, array items are consecutively stored in a column major order commonly used by math libraries written in FORTRAN such as LAPACK.

\subsection{Math Library Support}

As a sample scenario we implemented support for two important math libraries of common use. We wrote wrappers for LAPACK's singular value decomposition driver function \texttt{*gesvd} and the discrete Fourier transform functions of FFTW. These wrappers allow using the libraries directly inside SQL Server.

Since arrays a stored in exactly the same by other as required by the most common math libraries, and calling them only requires marshaling pointers between .NET and the native code, the overhead of these calls is negligible once the whole array is loaded into memory.

\section{Implementation Challenges}
\label{sec:implementation}

Since our application is somewhat different  from that SQL Server CLR is originally targeted to, we had to develop several tricks described in this section to implement the array library successfully and efficiently.

\subsection{Using C++/CLI}

We implemented the array library using C++/CLI 2008 to leverage some important features of the language that are not available in other .NET languages. An important objective of the array library was to provide array support to various base types (double, float etc.) which is very easy to implement using C++ templates but is not feasible using .NET generics due to the lack of pointer support (unsafe code) to generic types. Although C++/CLI has full support for .NET template classes it is tricky to write code that is interoperable with other .NET languages. When C++/CLI template classes are explicitly specialized   they are automatically exported into the binary library as strongly typed (i.e. non-generic) classes when the project is compiled.

The naming convention of the instantiated template classes, however, prevents using them from other .NET languages (at least from \csharp and Visual Basic). The C++/CLI compiles quoted class names and uses the special characters $<$ and $>$ in the quoted names but \csharp and VB do not support type name quotation. To overcome this drawback and benefit from the strength of C++/CLI we chose to do the following trick. The raw library compiled with C++/CLI containing the special class names was disassembled with the tools provided with the .NET SDK into IL assembly code. This code then contained all specially formatted class names as plain text and could be manipulated in a very simple way, special class names were replaced to unique simple names that are well-formatted for \csharp. We added this simple name replacement utility to our C++/CLI project as a final build step so the whole compilation process became transparent. One important drawback of this method is that the original debug symbols for C++/CLI could not be used any more and code had to be debugged using the symbols for the disassembled IL code.

\subsection{User-Defined Aggregates}
\label{sec:aggregates}

Although user-defined aggregate functions seem a very elegant way of implementing operations such as table to array conversion or covariance matrix computation due to various issues with the current SQL Server version, though we implemented the aggregates, we could not reach acceptable performance. The most important performance issue was that independently of the aggregate function internal storage requirements, the state of aggregation had to be serialized via a binary stream interface for each row processed by the aggregation. This turned out to be prohibitive in our scenarios.

In place of aggregate functions, we wrote plain SQL CLR scalar functions that take a SQL query as an input parameter of string, aggregate rows sequentially and return the resulting array as a binary blob.

\section{Library Interfaces}
\label{sec:interface}

\subsection{Manipulating Arrays from T-SQL}

Since we chose not to use UDTs to implement all array manipulation functions due to various performance issues the interface to the library became somewhat cumbersome. However, we organized functions under separate schemas by underlying data-type and storage class (see Sec.~\ref{sec:storage}) to eliminate extremely long function names. Functions acting on short (on-page) arrays of type INT are under the schema \texttt{IntArray}, the ones acting on max arrays (out-of-page) are under \texttt{IntArrayMax} etc. Functions accept and return arrays as \texttt{VARBINARY(8000)} or \texttt{VARBINARY(MAX)} depending on the storage class. Arrays are indexed with \texttt{SMALLINT} or \texttt{INT} values. Because Microsoft SQL Server does not support UDTs with a variable length of parameters many functions have numbered versions (denoted with an underscore and a number) accepting a certain number of parameters as we will show below.

An array of five floats, a vector of five elements, is created the following way:
\begin{lstlisting}
DECLARE @a VARBINARY(100) = 
     FloatArray.Vector_5(1.0, 2.0, 3.0, 4.0, 5.0)
\end{lstlisting}
To extract an item of this vector the following should be executed.
\begin{lstlisting}
SELECT FloatArray.Item_1(@a, 3)
\end{lstlisting}
This returns the third (zero indexed) element of the array. Higher dimensional arrays are accessed by the same function with more parameters, e.g.
\begin{lstlisting}
DECLARE @m VARBINARY(100) =
     FloatArray.Matrix_2(0.1, 0.2, 0.3, 0.4)
SELECT FloatArray.Item_2(@m, 1, 0)
\end{lstlisting}
Here the \texttt{Matrix\_2} function creates a 2-by-2 matrix from the listed four elements. \texttt{Item\_2} is used to retrieve items of two dimensional arrays.

Arrays can be created from row-by-row data stored in a table by two ways. However its performance is debatable (c.f. Sec.~\ref{sec:aggregates}), the \texttt{Concat} aggregate is used the following way:
\begin{lstlisting}
DECLARE @a VARBINARY(MAX)
DECLARE @l VARBINARY(100) =
     IntArray.Vector_2(100, 200)

SELECT @a = FloatArrayMax.Concat(@l, ix, v)
FROM table
\end{lstlisting}
Here the array is assembled from a table which has two columns: one containing the index of the item (as an array of two integers) and the value. The same functionality is implemented as a UDF accepting a T-SQL query as an input parameter and data is read by a \texttt{SqlDataReader} to retrieve the array elements. The latter method turned out to work much better than the obviously more elegant UDA.

Sub-arrays of an array can be retrieved using the \texttt{Subarray} function. The offset of the sub-array and the dimension sizes are the input parameters. Only retrieval of contiguous parts of the arrays is supported. The \texttt{Subarray} function is called the following way.
\begin{lstlisting}
DECLARE @a VARBINARY(MAX)
DECLARE @b VARBINARY(MAX)
...
SET @b = FloatArrayMax.Subarray(@a,
     IntArray.Vector_3(1, 4, 6),
     IntArray.Vector_3(5, 5, 5),
     0)
\end{lstlisting}
Here \texttt{@a} is initialized to be a three dimensional array of floats in the omitted code marked with the ellipsis. The parameters of the \texttt{SubArray} function are the array to subset, the offset of the subarray to retrieve, and the size of the subarray (here a 5-by-5-by-5 cube). The last parameter specifies whether subarrays with length of one in any dimension are automatically converted to a lower dimensional array. This is useful, for example, for retrieving the column vectors of a matrix.

Individual items of the arrays can be manipulated by the \texttt{UpdateItem} function:
\begin{lstlisting}
SET @a = FloatArray.UpdateItem_1(@a, 3, 4.5)
\end{lstlisting}

Arrays can be converted to tables by various table-valued functions, e.g. \texttt{ToTable}, \texttt{MatrixToTable} etc.

The function \texttt{Cast} is used to treat raw binaries containing consecutive numbers to be able to be treated as arrays by prefixing them with a header. The opposite to this is \texttt{Raw} which returns the array elements as a raw binary by stripping the header. The \texttt{Reshape} function is used to resize the array dimensions without reordering the array elements (original and target sizes must not differ). Conversion functions between different base types and storage classes exist. Arrays can also be converted to and from strings.

\subsection{Interface with .NET}

On the client-side arrays are visible as binary buffers or streams (containing the header) which have to be converted to .NET arrays first. The array library DLL can be referenced from any .NET language and contains functions for converting binaries to and from .NET arrays. The following \csharp code snippet uses a \texttt{SqlDataReader dr} to retrieve an array from the database:
\begin{lstlisting}
double[] v = dr.SqlFloatArray(dr.GetSqlBinary(1));
\end{lstlisting}
To convert a .NET array to SQL array it has to be converted to \texttt{SqlBinary} first. The SQL Server client for .NET can then be used to send the binary to save the binary into the database.
\begin{lstlisting}
double[] v = {1.0, 2.0, 3.0};
SqlFloatArray a = new SqlFloatArray(v);
SqlBinary x = a.ToSqlBuffer();
\end{lstlisting}

\subsection{Interface with Math Libraries}

Math libraries (to date only LAPACK and FFTW) are dynamically loaded by the array library and a T-SQL interface is provided via UDFs. For example, the Fourier transform on an array can be as easily performed from T-SQL as follows.
\begin{lstlisting}
DECLARE @ft VARBINARY(MAX)
SET @ft = FloatArrayMax.FFTForward(@a)
\end{lstlisting}
Here \texttt{@a} is an array stored in a \texttt{VARBINARY(MAX)} variable.
As the arrays are stored in a column major element order, interfacing with LAPACK is exceptionally easy, no transformation of the in-memory data is necessary, marshaling between .NET and the native code is simply done by reference. On the other hand, FFTW requires specially aligned memory buffers to perform well. When calling FFTW, a memory copy into a pre-aligned buffer is necessary but the performance gain is usually worth the otherwise expensive operation.

Additional third-party libraries can be added to the package by referencing the array library from any .NET language and implementing wrappers that connect the array data type with the custom library's interface and export the functionality as a Microsoft SQL Server UDF.

\section{Performance of the Library}
\label{sec:performance}

\subsection{Hardware and Software Configuration}

We ran the performance tests on a Dell PowerVault 2950 machine with two Intel Xeon Q8400 CPUs (four cores each) running at 2.67 GHz, 16 GB of RAM and an I/O subsystem yielding above 1~GB/s sequential read throughput for I/O limited scan operations. We used Microsoft Windows 2008 Server Enterprise Edition and Microsoft SQL Server 2008 Developer Edition for the test. Additional tests were conducted on a smaller machine running SQL Server 2008 R2 and the conclusions were not in contradiction with our results obtained with the older SQL Server version on the big machine so effectively there is no difference in the two versions with respect to user-defined function execution.

\subsection{Test Data}

We populated two tables with 357 million rows containing an ID (\texttt{Int64}, clustered index) and five dimensional vectors of double precision numbers. In the first table \texttt{Tscalar} the vector components were stored as scalar values in five separate columns \texttt{v1}, \texttt{v2}, ... \texttt{v5} while in the second table \texttt{Tvector} vectors were stored as a fix-sized binary column \texttt{v}. This second table had 24 bytes overhead per row resulting from the vector headers which made the whole table 43~\% bigger.

\subsection{The Test Queries}

The test queries were defined such a way that they would be executed as a simple clustered index scan operation reading all pages of the data table. The database server cache was explicitly cleared before each performance test run. The following queries were used.
\begin{enumerate}
\item \lstinline{SELECT COUNT(*) FROM Tscalar WITH (NOLOCK)}
\item \lstinline{SELECT COUNT(*) FROM Tvector WITH (NOLOCK)}
\item \lstinline{SELECT SUM(v1) FROM Tscalar WITH (NOLOCK)}
\item \lstinline{SELECT SUM(floatarray.Item_1(v, 0)) FROM Tvector WITH (NOLOCK)}
\item \lstinline{SELECT SUM(dbo.EmptyFunction(v, 0)) FROM Tvector WITH (NOLOCK)}
\end{enumerate}
Query~1 and~2 were used to compare the I/O cost of the clustered index scan operations while Query~3 and 4 were used to test the overhead coming from calls into the vector library extracting the first component of the vector. Query~5 was used for cross-checking with an empty function call to determine the overhead of the user-defined function calls. Table~\ref{tab:results} summarizes the metrics of our test queries.

\begin{table}
	\begin{center}
	\begin{tabular}{ | c | c | c | c |}
		\hline
Query & Execution time [s] & CPU load [\%] & I/O [MB/s] \\
		\hline
		\hline		
1	& 18	& 45	& 1150 \\
2	& 25	& 38	& 1150 \\
3	& 18	& 90	& 1150 \\
4	& 133	& 98	&  215 \\
5	& 109	& 99	&  265 \\
		\hline		
	\end{tabular}
	\caption{Query performance test results.}
	\label{tab:results}
	\end{center}
\end{table}

\section{Discussion}
\label{sec:discussion}

\subsection{Overhead of User-Defined Function Calls}

Table 1 shows that complex UDF function calls (small amount of data and frequent calls) easily lead to CPU-bound query performance. The cost of calling a CLR function for every row of the data table was 734~s (considering that Query~3 did not run at full CPU utilization and all eight cores were used). This yields a cost of about 2~$\mu$s per CLR function call. A detailed performance analysis revealed that at least 38~\% of the CPU time went for the UDF calls even when the UDF was empty.  When the array item was really extracted by the UDF (Query~4) the additional cost was 22~\% above the empty function call case. This cost came entirely from the managed functions written by us and very likely can be reduced by optimizing the code. 

\subsection{Extreme Scenario}

It is important to emphasize that we tested the worst-case scenario in which the relative overhead of the CLR UDF call is the highest compared to other costs of the operation. The tables we used were very narrow and the server I/O subsystem was very efficient, the database file layout was highly optimized for the capabilities of the RAID controllers. Also, the effective work done in the UDF was very limited. In production situations the relative overhead is not likely to be higher than our test results.

\section{Conclusions}
\label{sec:conclusions}

Implementing an array library for SQL Server using the CLR integration capabilities of the product is feasible but it is worth only if work done in the UDF is computationally intensive compared to the flat cost of calling the functions or the processing power of the hardware is way above the data rate provided by the I/O system. At the same time, the convenience of direct access to array functionality through T-SQL and the consistency of the programming interface to be used for sending data to external libraries is another major motivation for this approach.
Certain tiny updates to the UDT and UDA implementation of Microsoft SQL Server would make implementing a library like this much easier. The ability to serialize/deserialize small UDTs and UDAs from/to simple in memory byte buffers instead of using stream wrappers would allow implementing the whole library as member functions of the UDTs instead of the enormous number of individual functions. Also, the ability to partially serialize/deserialize large UDTs into the stream wrapping the B-tree containing out-of-page data would be necessary.

Although SQL CLR functions were designed to be simple enough to be used directly from T-SQL it is still somewhat cumbersome. A syntactic sugar to T-SQL and a pre-parser would be desirable that translates a special flavor of SQL designed for array notation to standard T-SQL with function calls. This could be achieved by writing a specialized .NET database connector that provides the translation.

\section{Acknowledgements}

The JHU group would like to thank the support of the Gordon and Betty Moore Foundation and Microsoft Research. ASz has been supported by the National Science Foundation grants NSF/AST-0428325 and NSF 09-530/0937947.

This work was supported by the following Hungarian grants: NKTH: Pol\'anyi, OTKA-80177 and KCKHA005.

\bibliographystyle{plain}
\bibliography{sqlarray}

\end{document}